\documentclass[aps,reprint,preprintnumbers,showpacs,showkeys,superscriptaddress]{revtex4-1}
\usepackage[T1]{fontenc}
\usepackage{amsfonts,amsmath,amssymb,amsbsy}
\usepackage{graphicx,color}
\usepackage{times}
\usepackage{upgreek}
\usepackage{bbold} 
\definecolor{yblue}{rgb}{0.06, 0.3, 0.57}
\usepackage[colorlinks=true, linkcolor=yblue, citecolor=yblue, urlcolor=yblue]{hyperref}
\let\mathbf=\boldsymbol

\begin{document}

\title{Motion of skyrmions in nanowires driven by magnonic momentum-transfer forces}

\author{Xichao Zhang}
\thanks{These authors contributed equally to this work.}
\affiliation{School of Science and Engineering, The Chinese University of Hong Kong, Shenzhen 518172, China}
\affiliation{School of Electronic Science and Engineering, Nanjing University, Nanjing 210093, China}

\author{Jan M\"uller}
\thanks{These authors contributed equally to this work.}
\affiliation{Institut f\"ur Theoretische Physik, Universit\"at zu K\"oln, D-50937 Cologne, Germany}

\author{Jing Xia}
\affiliation{School of Science and Engineering, The Chinese University of Hong Kong, Shenzhen 518172, China}

\author{Markus Garst}
\email[E-mail:~]{markus.garst@tu-dresden.de}
\affiliation{Institut f\"ur Theoretische Physik, Technische Universit\"at Dresden, D-01062 Dresden, Germany}

\author{Xiaoxi Liu}
\affiliation{Department of Information Engineering, Shinshu University, 4-17-1 Wakasato, Nagano 380-8553, Japan}

\author{Yan Zhou}
\email[E-mail:~]{zhouyan@cuhk.edu.cn}
\affiliation{School of Science and Engineering, The Chinese University of Hong Kong, Shenzhen 518172, China}

\begin{abstract}
We study the motion of magnetic skyrmions in a nanowire induced by a spin-wave current $J$ flowing out of a driving layer close to the edge of the wire. By applying micromagnetic simulation and an analysis of the effective Thiele equation, we find that the skyrmion trajectory is governed by an interplay of both forces due to the magnon current and the wire boundary. The skyrmion is attracted to the driving layer and is accelerated by the repulsive force due to the wire boundary. We consider both cases of a driving longitudinal and transverse to the nanowire, but a steady-state motion of the skyrmion is only obtained for a transverse magnon current. For the latter case, we find in the limit of low current densities $J$ the velocity-current relation $v \sim J/\alpha$ where $v$ is the skyrmion velocity and $\alpha$ is the Gilbert damping. For large $J$ in case of strong driving, the skyrmion is pushed into the driving layer resulting in a drop of the skyrmion velocity and, eventually, the destruction of the skyrmion.
\end{abstract}

\date{June 5, 2017}
\preprint{\href{http://iopscience.iop.org/article/10.1088/1367-2630/aa6b70}{\textit{New J. Phys.} \textbf{19}, 065001 (2017)}}

\maketitle

\section{Introduction}
\label{se:introduction}

Magnetic skyrmions are topological protected textures of the magnetization~\cite{Roszler_NATURE2006,Nagaosa_NNANO2013,Seki_BOOK2016}, that have been observed experimentally in various magnetic materials, in bulk systems, thin magnetic films, and magnetic multilayers~\cite{Muhlbauer_SCIENCE2009,Yu_NATURE2010,Heinze_NPHYS2011,Schulz_NPHYS2012,Romming_SCIENCE2013,Finazzi_PRL2013,Kezsmarki_NMATER2015,Schwarze_NMATER2015,Du_NCOMMS2015,Nii_NCOMMS2015,Wanjun_SCIENCE2015,Boulle_NNANO2016,Wanjun_NPHYS2017,Wanjun_AIP2016,WOO_NMAT2016}. As a topological object, the skyrmion is stable and possesses a peculiar particle-like nature, which makes it suitable for the application as an information carrier. Indeed, a number of theoretical and numerical works have demonstrated that magnetic skyrmions could be essential components for future magnetic and spintronic devices for data storage and computation~\cite{Fert_NNANO2013,Sampaio_NNANO2013,Iwasaki_NNANO2013,Sun_PRL2013,Tomasello_SREP2014,Yan_NCOMMS2014,Xichao_SREP2015A,Xichao_SREP2015B,Xichao_SREP2015C,Upadhyaya_PRB2015,Koshibae_JJAP2015,Yan_NCOMMS2015,Fusheng_NANOLETT2015,Beg_SREP2015,Xichao_NCOMMS2016,Mueller_ARXIV2016,Wang_SREP2016,Senfu_NJP2015,Beg_PRB2017,Bazeia_JMMM2017}.

The position of an isolated magnetic skyrmion can be manipulated by an external driving force. A spin-polarized electric current has been reported to be an effective driving force for the motion of magnetic skyrmions in confined geometries~\cite{Sampaio_NNANO2013,Iwasaki_NNANO2013,Wanjun_SCIENCE2015,Lin_JAP2014,Reichhardt_PRB2015,Guoqiang_NL2017}. In addition, the scattering of a propagating spin wave has been demonstrated to generate a momentum-transfer resulting in a skyrmion or domain wall motion~\cite{Schutte_PRB2014A,Schutte_PRB2014B,Iwasaki_PRB2014,Schroeter_LTP2015,Xichao_NANOTECH2015,Kim_ARXIV2015,Wanjun_PRL2013,Wang_PRL2011,Wang_PRL2012}. In the absence of boundaries, the skyrmion will be driven towards the magnon source, i.e., the skyrmion velocity posses a component antiparallel to the magnon current~\cite{Schutte_PRB2014A,Schutte_PRB2014B,Iwasaki_PRB2014,Schroeter_LTP2015}. In confined geometries, the boundary acts as a potential barrier and its repulsive force can also result in an effective motion parallel to the magnon current~\cite{Xichao_NANOTECH2015,Kim_ARXIV2015}.

In this paper, we study in detail the motion and dynamics of an isolated magnetic skyrmion in a magnetic nanowire driven by spin waves travelling longitudinal or transverse to the wire. As the magnon current decays on a length scale set by the Gilbert damping, the longitudinal driving is only viable for short wires. For transverse driving, a steady-state skyrmion motion is obtained with a characteristic velocity-current relation. We find that it is determined by an interplay of the magnonic driving and the repulsive potential arising from the edge of the wire. For large magnon currents, however, the skyrmion is pushed into the driving layer, that generates the spin waves, leading to a breakdown of the skyrmion velocity. Our results provide a guide for future experimental studies on skyrmion motion in confined geometries driven by magnonic momentum-transfer forces.

\section{Results}
\label{sec:results}

We consider a magnetic wire with a surface-induced Dzyaloshinskii-Moriya interaction that stabilizes magnetic skyrmions, see Appendix~\ref{sec:methods_1} for details. The motion of an isolated magnetic skyrmion is investigated that is driven by spin waves propagating longitudinal and transverse to the wire. We first discuss the resulting skyrmion trajectories and then turn to a discussion of the relation between the driving current and the skyrmion velocity in the steady state.

\subsection{Skyrmion trajectories for longitudinal and transverse driving by spin waves}

\begin{figure*}[t]
  \centerline{\includegraphics[width=1.00\textwidth]{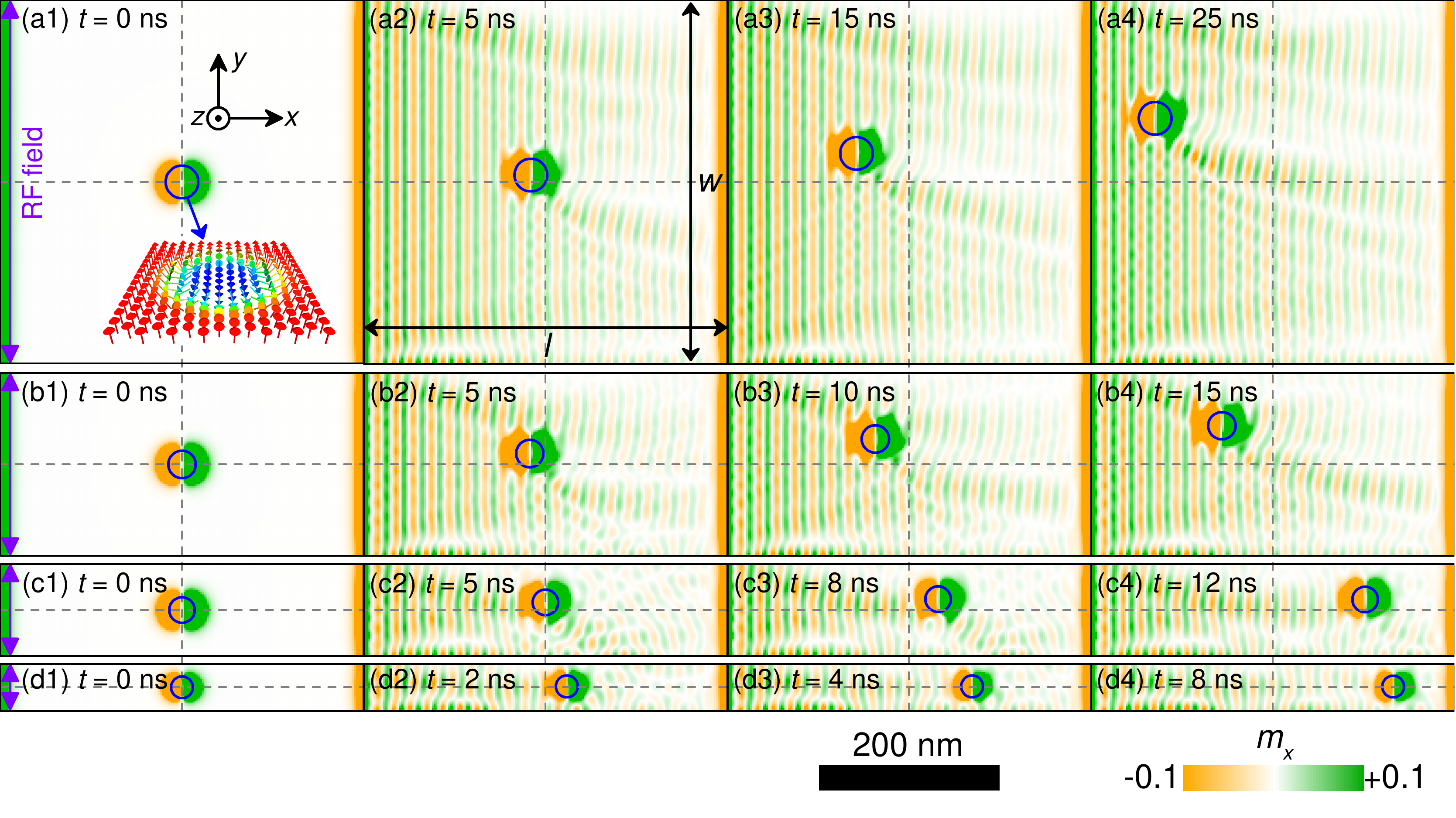}}
  \caption{%
    Snapshots of a skyrmion driven by magnons parallel to the nanostrip of length $l=400$ nm. 
    (a1)-(a4): the width $w=400$ nm of the strip is equal to the length; 
    (b1)-(b4): $w=200$ nm; 
    (c1)-(c4): $w=100$ nm; 
    (d1)-(d4): $w=50$ nm; 
    time steps of the snapshots are indicated. 
    An RF field with an amplitude of $H_{\text{a}}=1000$ mT and a frequency of $f=200$ GHz is applied at the left edge of all samples, producing magnons traveling towards the right edge. 
    The Gilbert damping $\alpha=0.02$. 
    The inset of (a1) shows the structure of the N\'eel-type skyrmion in our simulation, which is indicated by blue circles in all snapshots. 
    The color scale shows the in-plane component of the magnetization $m_x$, which is rescaled to $[-0.1,+0.1]$ in order to show the magnon profile more clearly. 
    A length scale is also provided.
  }
  \label{FIG1}
\end{figure*}

\begin{figure*}[t]
  \centerline{\includegraphics[width=1.00\textwidth]{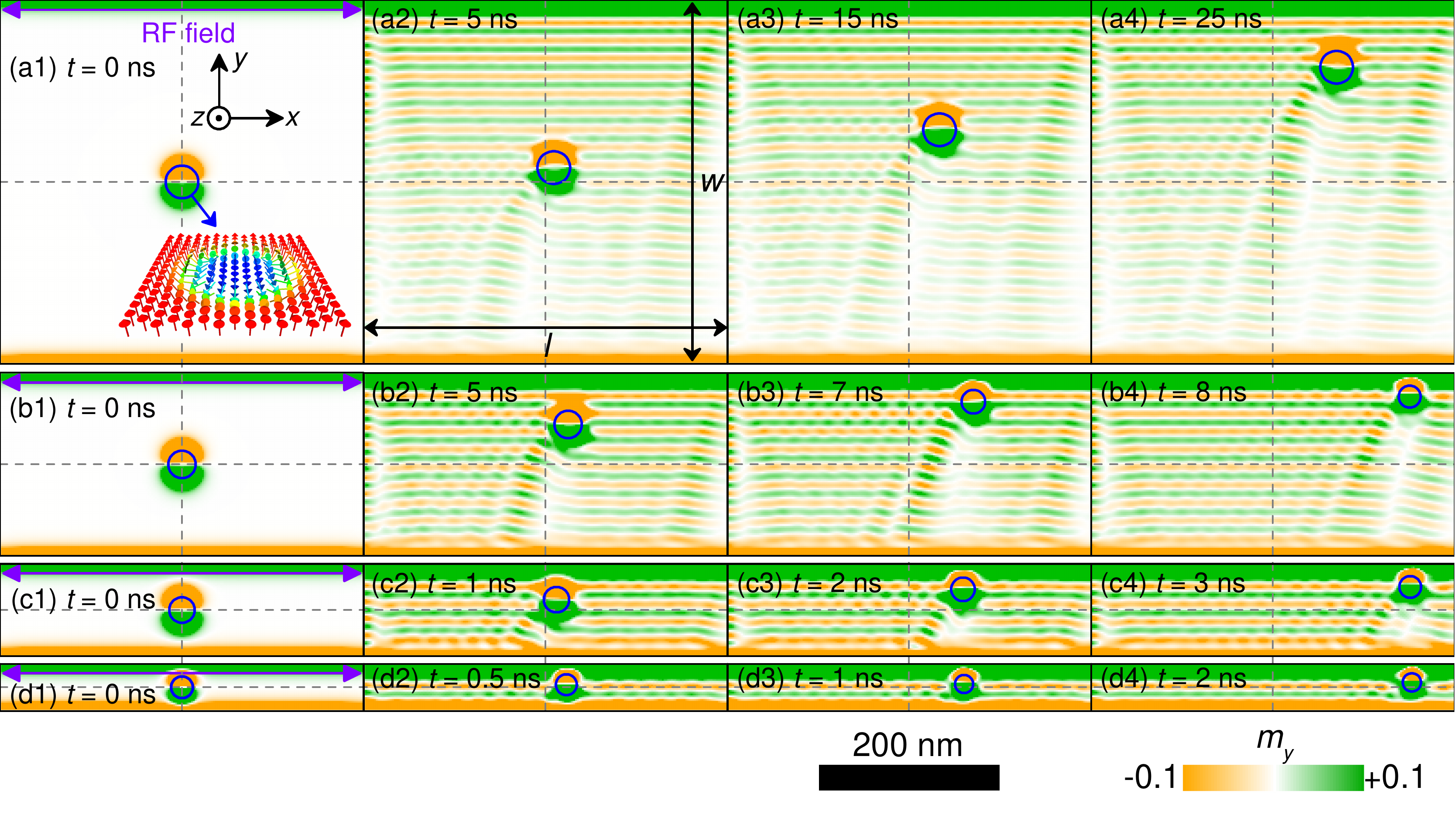}}
  \caption{%
    Snapshots of a skyrmion driven by magnons transverse to the nanostrip of length $l=400$ nm. 
    (a1)-(a4): the width $w=400$ nm of the strip is equal to the length; 
    (b1)-(b4): $w=200$ nm;
    (c1)-(c4): $w=100$ nm;
    (d1)-(d4): $w=50$ nm;
    time steps of the snapshots are indicated. 
    An RF field with an amplitude of $H_{\text{a}}=1000$ mT and a frequency of $f=200$ GHz is applied at the lower edge of all samples, producing magnons traveling towards the upper edge. 
    The Gilbert damping $\alpha=0.02$. 
    The inset of (a1) shows the structure of the N\'eel-type skyrmion in our simulation, which is indicated by blue circles in all snapshots. 
    The color scale shows the in-plane component of the magnetization $m_y$, which is rescaled to $[-0.1,+0.1]$ in order to show the magnon profile more clearly. 
    A length scale is also provided.
  }
  \label{FIG2}
\end{figure*}

We first present results for the skyrmion trajectories obtained with the help of micromagnetic simulations. As shown in Figs.~\ref{FIG1} and \ref{FIG2}, we consider a nanowire consisting of a magnetic layer with a length $l = 400$ nm in the $x$-direction and various different widths $w = 400, 200, 100,$ and $50$ nm in the $y$-direction corresponding to panels (a)-(d), respectively. The thickness of the magnetic layer in the $z$-direction is fixed at $1$ nm. The initial magnetization profile of the magnetic layer corresponds to a magnetization pointing along the $+z$-direction except at the center of the sample, where the skyrmion is initially located, and at the sample edges, where the magnetization is tilted due to the Dzyaloshinskii-Moriya interaction.

Two setups with longitudinal and transverse driving are considered, Figs.~\ref{FIG1} and \ref{FIG2}, respectively. The driving is generated by a locally applied oscillating magnetic field, that is, a radio frequency (RF) field. It is applied only within a narrow strip of width $15$ nm that is either located on the left-hand side of the sample for longitudinal driving or at the top of the nanowire for transverse driving. We consider a RF field $\boldsymbol{H}=H_{\text{a}}\sin{(2\pi ft)}\hat{s}$ with amplitude $H_{\text{a}}=1000$ mT and frequency $f=200$ GHz. It possesses a longitudinal polarization with $\hat s = \hat y$ for Fig.~\ref{FIG1} and $\hat s = \hat x$ for Fig.~\ref{FIG2}. More details on the simulation and material parameters are given in Appendix~\ref{sec:methods_1}.

We first discuss the situation of {\it longitudinal driving}. Figure~\ref{FIG1}(a1) shows the case of a square-shaped thin film where $l=w=400$ nm at time $t=0$. The panels (a2)-(a4) show snapshots at later times after the driving field has been switched on. The RF field at the left edge produces spin waves traveling towards the right, along the length direction of the sample. Since the magnetic skyrmion is far away from the sample edges, we observe a nearly pure skew scattering between the propagating spin wave and the magnetic skyrmion, which is in good agreement with that reported in Ref.~\onlinecite{Iwasaki_PRB2014}. Indeed, the spin wave-skyrmion scattering leads to a backwards motion of the magnetic skyrmion. It can be seen that the magnetic skyrmion basically moves against the propagation direction of the spin wave, and reaches the left edge in a finite time. In addition, the skyrmion gets slightly dragged towards the upper edge.

The width in Figs.~\ref{FIG1}(a) is sufficiently large so that the skyrmion reaches the magnon source before it touches the upper edge of the wire. The situation changes when the track is narrower. In Figs.~\ref{FIG1}(b) the width is $w = 200$ nm and the skyrmion still has not reached the upper edge at a time $t = 15$ ns. However, for even narrower wires with $w = 100$ nm and $w = 50$ nm, see Figs.~\ref{FIG1}(c) and \ref{FIG1}(d), it reaches the edge after a short time. Close to the edge, the skyrmion changes the direction of its motion. Instead of approaching the driving layer, the skyrmion moves along the edge away from it. This evading motion along the edge is also much faster than the attractive motion towards the magnon source.

For the {\it transverse driving} we observe similar effects. The RF field is applied on the upper edge of the nanowire, see Fig.~\ref{FIG2}, in the same wire geometries as for the parallel driving. Due to the oscillating magnetic field, magnons are excited and propagate downwards along the $-y$-direction. The interaction with magnons pushes the skyrmion towards the upper edge of the wire with a slight side-shift to the right-hand side. This can be particularly well seen for the wide wire in Figs.~\ref{FIG2}(a). This setup is however too wide to observe the effects from the sample edge within the simulated time span. In contrast to the longitudinal driving mechanism, the skyrmion is now driven faster towards the upper edge, which becomes apparent for the narrower wires shown in Figs.~\ref{FIG1}(b)-(d). When the skyrmion arrives at the edge, it speeds up dramatically and moves along the upper edge towards the right-hand side. It reaches the end of the wire in a much shorter time than for longitudinal driving.

\begin{figure}[t]
  \centerline{\includegraphics[width=0.47\textwidth]{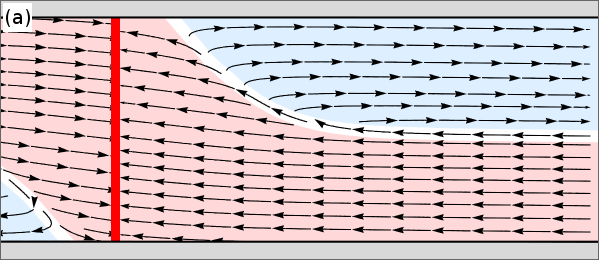}}
  \vspace{1mm}
  \centerline{\includegraphics[width=0.47\textwidth]{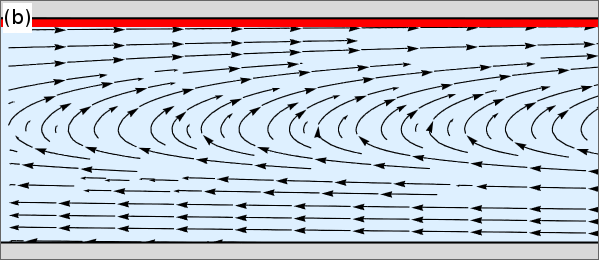}}
  \caption{%
    Trajectories (black arrows) of a magnetic skyrmion derived within the Thiele approximation. The skyrmion is driven by magnons from a source (red line) and flowing 
    longitudinal (a) and transverse (b) to the nanotrack (gray boundaries).
    Skyrmions in one of the colored areas either run free along the track (blue) or get captured by the magnon source (red).
    Areas with different scenarios are divided by a separatrix (white).
    Parameters are $D/|G|=1$, $\alpha=0.01$, $\sigma_{\parallel}/L_\text{edge}=0.1$, $\sigma_{\perp}/L_\text{edge}=0.1$, $w/L_\text{mag}=5$ for both plots and
    $L_\text{mag}/L_\text{edge} = 1$ and $J q/V_\text{edge} = 3$ for the parallel driving (a) and 
    $L_\text{mag}/L_\text{edge} = 3$ and $J q/V_\text{edge} = 0.01$ for the transverse driving (b).
  }
  \label{FIG3}
\end{figure}

\subsection{Thiele approximation}

The skyrmion trajectories observed by our micromagnetic simulation can be understood in the framework of an effective Thiele equation of motion for the skyrmion~\cite{Thiele_PRL1973}. Within this description, the skyrmion coordinate $\mathbf{R}$ is governed by the equation
\begin{equation}
\mathbf{G}\times \mathbf{\dot{R}} + \alpha \mathcal{D} \cdot \mathbf{\dot{R}} = \mathbf{F}(\mathbf{R})
\label{eq:thiele}
\end{equation}
where $\mathbf{G}= G\hat z$ is the gyrocoupling vector with $G<0$, which is related to the topological winding number of the skyrmion, $\alpha$ is the Gilbert damping, and $\mathcal{D}$ the dissipative tensor which can be approximated to be diagonal, $\mathcal{D}_{ij} = \mathcal{D} \delta_{ij}$. The force on the right-hand side is attributed to the magnon driving and the edge of the sample, $\mathbf{F} = \mathbf{F}_\text{edge} + \mathbf{F}_{\rm mag}$.

Close to the edge, the  Dzyaloshinskii-Moriya interaction leads to a twist of the magnetization~\cite{Rohart_PRB2013} that acts as a repulsive potential for the skyrmion~\cite{Iwasaki_NNANO2013,Xichao_SREP2015A}. To a good approximation, this potential falls off exponentially with the distance to the edge~\cite{Navau2016}. We thus describe the repulsive force by the edges of a nanostrip of width $w$ by a superposition of edges at positions $y_1=0$ and $y_2=w$:
\begin{equation}
\mathbf{F}_\text{edge}(\mathbf{R})= -V_\text{edge} \nabla \left( e^{-\frac{y}{L_\text{edge}}} + e^{\frac{y-w}{L_\text{edge}}} \right) \text{.}
\end{equation}
where $V_\text{edge} > 0$ parametrizes the strength of the potential and $L_\text{edge}$ the penetration depth of the magnetization twist.

The momentum-transfer from the magnon current to the skyrmion also results in a force,
\begin{equation}
\mathbf{F}_\text{mag}(\mathbf{R}) = J e^{-\frac{\mathbf{r}\cdot\hat{q}}{L_\text{mag}}} q \left( \sigma_{\parallel} \hat{q} + \sigma_{\perp} (\hat{z} \times \hat q) \right),
\end{equation}
where $J > 0$ is the two-dimensional magnon-current density, and $\vec q$ is the wavevector of the spin wave with $q = |\vec q|$ and $\hat q = \vec q/q$. It was shown in Refs.~\cite{Schutte_PRB2014A} that the force is determined by the two-dimensional transport scattering cross section of the skyrmion, $\sigma_{\parallel}$ and $\sigma_{\perp}$, longitudinal and transverse to the flow direction $\hat q$ of the magnon current. In general, they depend in a non-trivial manner on the magnon frequency. In the high-frequency limit or, equivalently, for large magnon wavevectors $q$ \cite{Schroeter_LTP2015}, the transverse transport scattering cross section is universal $\sigma_{\perp} \approx 4\pi/q$ and $\sigma_{\parallel} \sim 1/q^2$ so that $\sigma_{\perp} > \sigma_{\parallel} > 0$. We also accounted for the decay of the magnon current on a length scale set by the Gilbert damping, $1/L_{\rm mag} \approx \alpha \sqrt{\frac{m}{2\hbar} 2\pi f}$ where $m$ is the magnon mass and $f$ is the frequency of the wave.

The solution of the Thiele equation is plotted in Fig.~\ref{FIG3} for some set of parameters. For longitudinal driving in panel (a), the wire can be divided into two different areas. The skyrmion trajectories belonging to the red shaded area, on the one hand, will end up at the driving layer (red line).
When the skyrmion starts within the blue shaded area, on the other hand, it will be driven away from the driving layer. The interplay between the magnon and edge forces dominates the motion. As the magnon current decays exponentially on the length scale $L_\text{mag}$ with increasing distance to the driving layer, the magnon force that keeps the skyrmion close to the edge also decays so that eventually the skyrmion slowly approaches the center of the wire. The red and blue shaded areas are separated by a critical skyrmion trajectory (white line). In any case, the longitudinal driving setup will not produce a steady state.

In case of transverse driving in Fig.~\ref{FIG3}(b), a skyrmion initially positioned at the center of the wire gets attracted towards the driving layer at the top of the wire. At the same time, it gets repelled by the edge twist of the magnetization and if the driving is not too strong the skyrmion reaches a steady state with $v_{ys} = 0$ and a constant velocity, $v_{xs}$, along the edge. Within the Thiele approximation, this saturated velocity is given by
\begin{align} \label{Scaling}
v_{xs} = \frac{J}{\alpha D} e^{y/L_\text{mag}} q \sigma_\perp \approx \frac{4\pi J}{\alpha D}
\end{align}
It depends on the steady state distance to the edge, $y$, which is in turn governed by the driving amplitude $J$. The last equation applies in the high-frequency limit $\sigma_{\perp} \approx 4\pi/q$ and, in addition, $y/L_\text{mag} \ll 1$. In the limit of a small driving amplitude $H_\text{a}$, the current $J$ will obey Fermi's Golden rule $J \propto |H_\text{a}|^2$. In this case, the saturated velocity will scale as $v_{xs} \propto |H_\text{a}|^2/\alpha$.

\subsection{Damping dependence of the skyrmion motion}

\begin{figure*}[t]
  \includegraphics[width=1.00\textwidth]{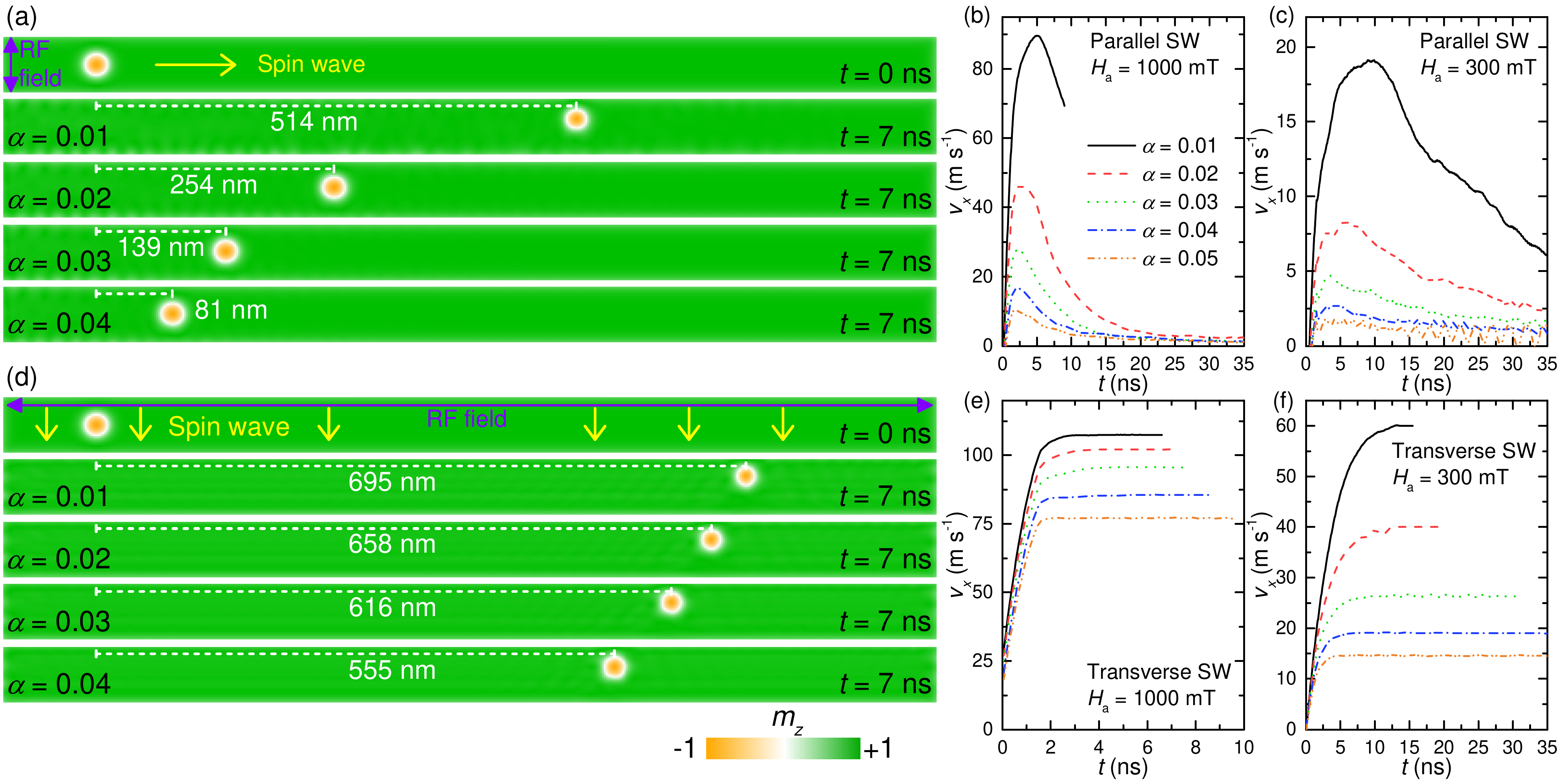}
  \caption{%
    Travelled distance and velocity of a single skyrmion in a nanowire ($1000$ nm $\times$ $60$ nm $\times$ $1$ nm).
    Snapshots of the system after $7$ ns are shown for longitudinal (a) and transverse (d) driving with different damping coefficients $\alpha$.
    Magnons are excited by an RF field as indicated with $H_{\text{a}}=1000$ mT and $f=180$ GHz. 
    The color scale represents the out-of-plane component of the magnetization $m_z$.
    In panel (b), (c) and (e),(f), the velocity parallel to the wire as a function of time, $v_x(t)$, is shown for longitudinal and transverse driving for two different driving amplitudes 
    $H_{\text{a}}=1000$ mT  and $H_{\text{a}}=300$ mT and for various damping coefficients $\alpha$ as given in the inset of (b).
    The curves are only plotted in the regime, where the skyrmion is sufficiently far from the end of the wire.
  }
  \label{FIG4}
\end{figure*}

In order to study the dependence of the skyrmion motion on the Gilbert damping parameter $\alpha$, we performed further simulations with a long and narrow wire with $w = 60$ nm and $l = 1000$ nm for various values of $\alpha$ in the range from $0.01$ to $0.05$. We consider two amplitudes for the excitation field  $H_{\text{a}}=300$ mT and $H_{\text{a}}=1000$ mT with a frequency $f=180$ GHz. The results are shown in Fig.~\ref{FIG4}. All setups share the common property, that the skyrmion moves faster and also further if the damping is lower. The exact dependence on the damping is, however, very different for longitudinal and transverse driving.

For longitudinal driving in Fig.~\ref{FIG4}(a), the skyrmion started within the blue shaded area of Fig.~\ref{FIG3}(a) so that its motion is eventually along the edge of the wire. The distance travelled after a time of $7$ ns for a driving amplitude $H_{\text{a}}=1000$ mT decreases approximately exponentially with increasing $\alpha$. While for the lowest simulated damping, $\alpha=0.01$, the skyrmion travels $514$ nm, it travels only half as far ($254$ nm) if the damping is doubled to $\alpha=0.02$. If we increase the damping further by another $\Delta\alpha=0.01$, the travelled distance again decreases by a factor $2$ to only $139$ nm. At even larger damping, $\alpha=0.04$, the skyrmion moves $81$ nm.
The corresponding components of the velocity along the track, $v_x(t)$, are shown in Fig.~\ref{FIG4}(b). The period of acceleration lasts approximately $5$ ns for low damping, $\alpha=0.01$, and becomes shorter with increasing damping. After the peak velocity is reached, the skyrmion decelerates again. This peak velocity can be up to $\sim 90$ m s$^{-1}$ for $\alpha=0.01$, while it is only $\sim 45$ m s$^{-1}$ for $\alpha=0.02$ and $\sim 25$ m s$^{-1}$ for $\alpha=0.03$. If we lower the RF field amplitude down to $H_{\text{a}}=300$ mT, see Fig.~\ref{FIG4}(c), the overall shape of the velocity curves stays the same but the time axis rescales with a factor $\sim 2$ while the value of the velocity rescales by $\sim 0.2$. The highest velocity, at $\alpha=0.01$, is now only $\sim 20$ m s$^{-1}$ and already $< 5$ m s$^{-1}$ for $\alpha=0.03$. For $\alpha>0.03$ the motion is almost immediately damped out.

For transverse driving in Fig.~\ref{FIG4}(d), the skyrmion moves much faster and further as compared to longitudinal driving. For low damping $\alpha=0.01$, the skyrmion travels $695$ nm to be compared with $514$ nm for longitudinal driving. For large damping $\alpha=0.04$ the skyrmion still travels $555$ nm. A closer look at the corresponding velocity, see Fig.~\ref{FIG4}(e), reveals that after an acceleration time of $\sim 2$ ns the skyrmion velocity reaches a steady state.
For all simulated damping constants, $\alpha \in [0.01, 0.05]$, the saturated velocities, $v_{xs}$, are in the regime of $75$ up to $110$ m s$^{-1}$ for $H_{\text{a}}=1000$ mT, which is only comparable to the peak velocity at lowest damping for longitudinal driving. The dependence of $v_{xs}$ on the Gilbert damping becomes much more pronounced for smaller driving amplitude $H_{\text{a}}=300$ mT as shown in Fig.~\ref{FIG4}(f).

\subsection{Steady-state motion for transverse driving}

\begin{figure}[t]
  \centerline{\includegraphics[width=0.47\textwidth]{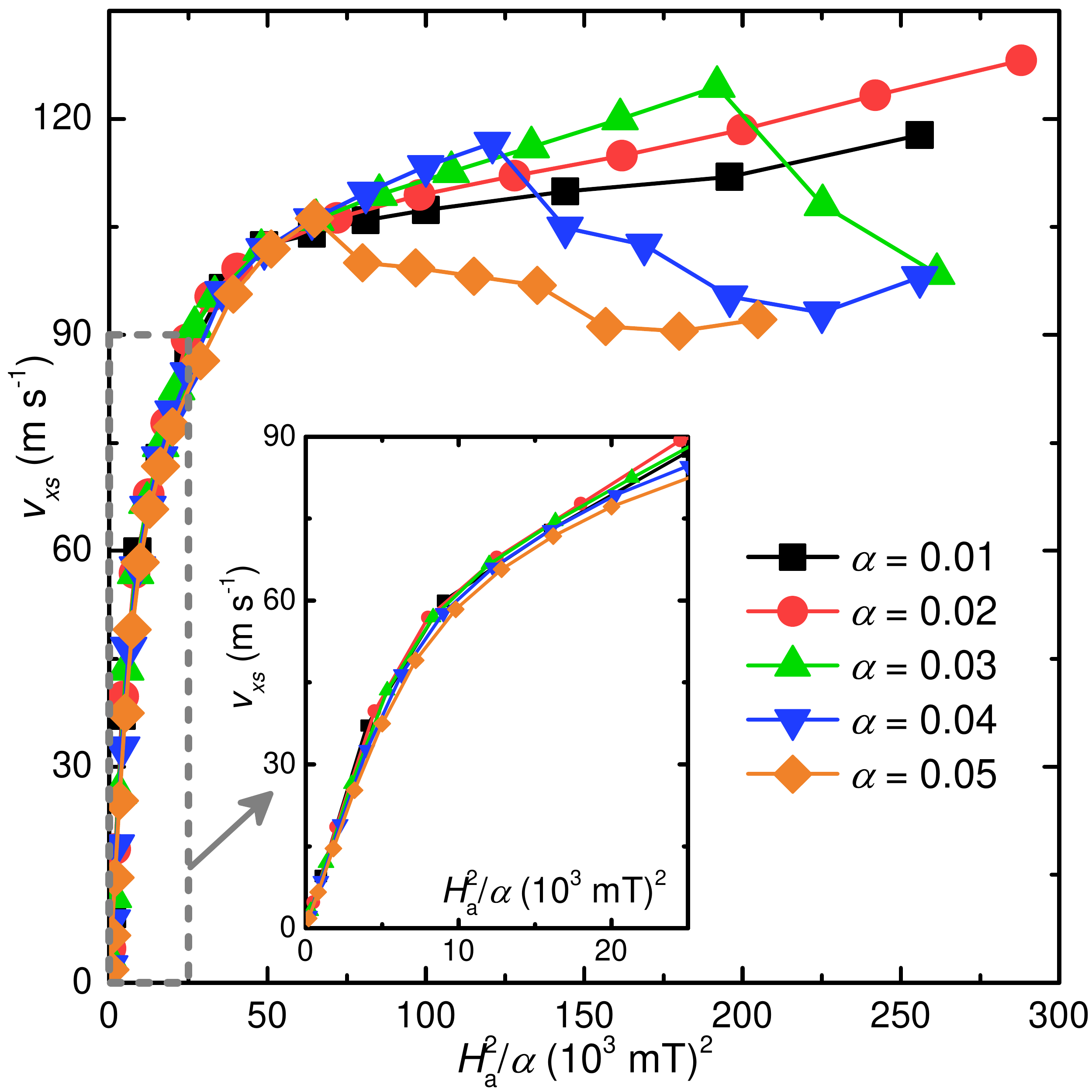}}
  \caption{%
    Saturation velocity $v_{xs}$ of a skyrmion in a nanostrip as a function of the scaling variable $H_\text{a}^2/\alpha$ where $H_\text{a}$ is the driving amplitude and $\alpha$ is the Gilbert damping. Results from micromagnetic simulations for transverse driving. The stimulating field $H_{\text{a}}$ ranges from $0$ to $3200$ mT and the frequency is fixed to $f=180$ GHz. Dimensions of the strip are $1000$ nm $\times$ $60$ nm $\times$ $1$ nm.
    The saturation velocity $v_{xs}$ is measured when the skyrmion velocity is converged. 
    The inset shows a close-up for low driving where the data approximately collapses onto one single curve.
  }
  \label{FIG5}
\end{figure}

For transverse driving, the skyrmion motion can assume a steady state with a constant, saturated velocity. In order to compare with the expression in Eq.~\eqref{Scaling} predicted by the Thiele approximation, we obtained the saturated velocity for transverse driving with the help of micromagnetic simulations using damping coefficients $\alpha = 0.01$ up to $0.05$ and driving amplitudes within the range from $0$ to $3200$ mT at a fixed frequency of $180$ GHz. The results are shown in Fig.~\ref{FIG5} as a function of the scaling variable $H_{\text{a}}^2/\alpha$.

For small values of $H_{\text{a}}^2/\alpha < 10\times \left(10^{3}~\mathrm{mT}\right)^2$, the data collapses onto a universal straight line corroborating the prediction from the Thiele approximation~\eqref{Scaling} using Fermi's Golden rule $J \propto |H_\text{a}|^2$. For larger values, the data still follows approximately a common curve but the steep increase of $v_{xs}$ flattens out. Close to the value $H_{\text{a}}^2/\alpha \approx 70\times \left(10^{3}~\mathrm{mT}\right)^2$, the saturated velocity $v_{xs}$ with damping $\alpha = 0.05$ shows a cusp and abruptly decreases. Subsequently, at higher values of $H_{\text{a}}^2/\alpha$ the velocities for smaller damping parameters also exhibit a sudden breakdown.

\begin{figure}[t]
  \centerline{\includegraphics[width=0.50\textwidth]{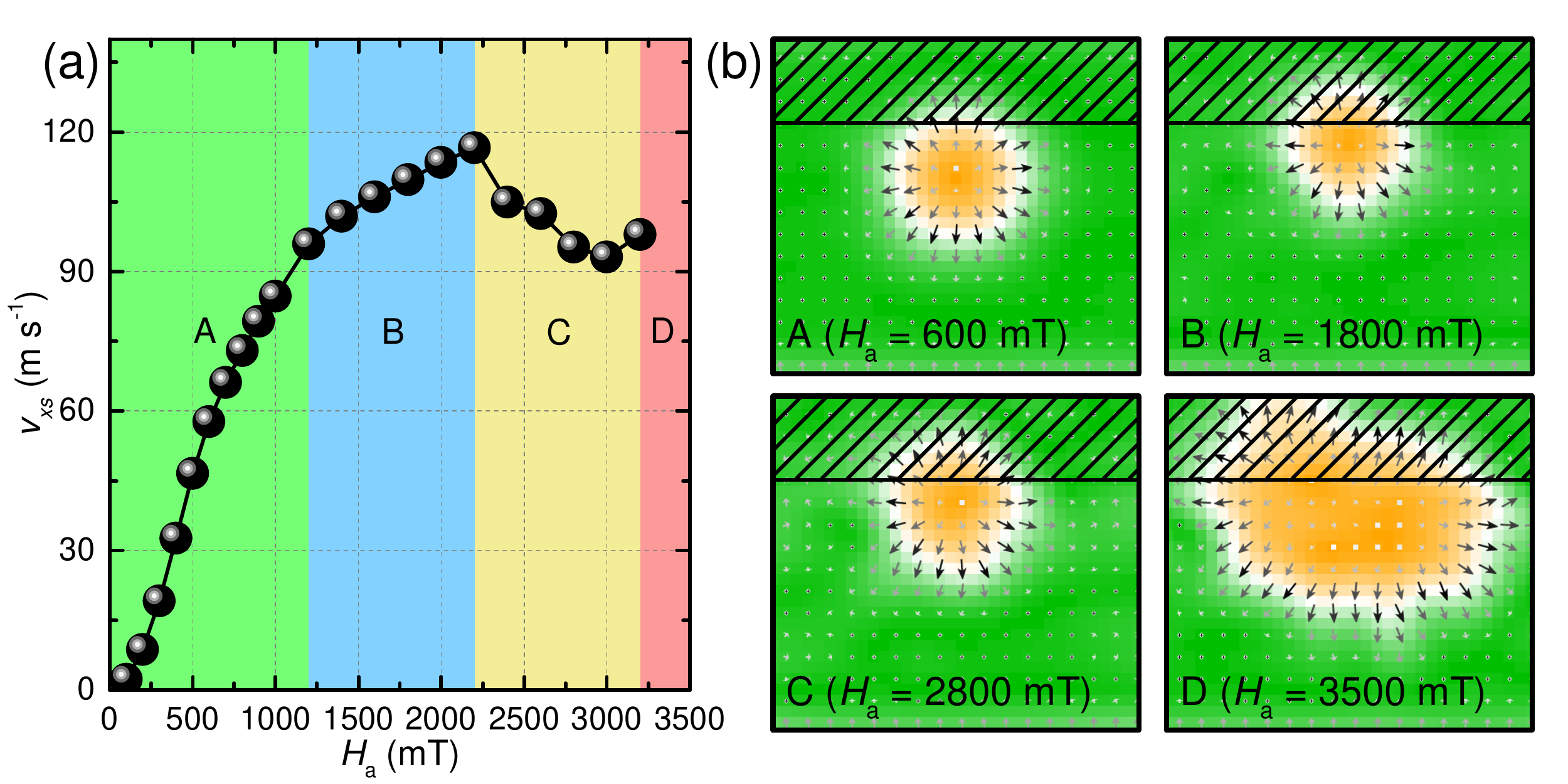}}
  \caption{%
    (a) The saturation velocity $v_{xs}$ of a skyrmion driven by a transverse magnon current in a nanostrip as a function of the stimulating field amplitude $H_\text{a}$. 
    Results from micromagnetic simulations.
    Four different scenarios are marked with a letter and background color.
    ``A'', green: the skyrmion moves rigidly along the edge of the nanotrack.
    ``B'', blue: the skyrmion moves along the nanotrack and simultaneously emits spin waves. 
    ``C'', yellow: the skyrmion moves along the nanotrack, emits spin waves, and periodically expands/shrinks in size, i.e., its {\it breathing mode} is excited. 
    ``D'', red: the skyrmion is driven into the edge and gets destroyed when touching it. 
    The frequency of the RF field is $f=180$ GHz and the damping coefficient is $\alpha=0.04$.
    Dimensions of the strip are $1000$ nm $\times$ $60$ nm $\times$ $1$ nm.
    (b)
    Snapshots of the magnetization in the four regimes described in (a).
    The RF field is applied at the upper edge with a width of $15$ nm, which is indicated by the shadowed area. 
    The color scale shows the component of the magnetization $m_z$, perpendicular to the film.
    Black arrows indicate the in-plane components.
  }
  \label{FIG6}
\end{figure}

In order to elucidate this sudden decrease of the saturated velocity, we show in Fig.~\ref{FIG6}(a) the saturated velocity $v_{xs}$ versus the driving amplitude $H_\text{a}$ at a fixed damping $\alpha = 0.04$. Depending on the strength of $H_\text{a}$, we can distinguish four different scenarios A to D. Snapshots of the skyrmion corresponding to these scenarios are shown in panel (b). The shaded area at the top corresponds to the strip where the oscillating driving field is applied. The scenarios are characterized as follows. 

Scenario A (green region) corresponds to lowest driving fields. In the range from $H_{\text{a}} = 0$ up to $1200$ mT, the saturated velocity $v_{xs}$ first increases with $H_{\text{a}}^2/\alpha$ as predicted from the Thiele approximation and then turns to a less steep increase. In this regime, the skyrmion smoothly moves along the nanotrack, and does not suffer any significant deformations. Moreover, only a minor fraction of the skyrmion area has entered the region where the oscillating RF field is applied, see Fig.~\ref{FIG6}(b-A).

Scenario B (blue region) is the intermediate regime between $H_{\text{a}}=1200$ and $2200$ mT. Here, the saturated velocity still increases with the amplitude of the stimulating RF field but the increase is less steep than in scenario A. This is related to a power loss from the emission of additional spin waves by the skyrmion, which can be discerned in our simulations.

Scenario C (yellow region) is obtained in the range from $H_{\text{a}}=2200$ up to $3200$ mT which is the highest accessible driving amplitude. In this regime the saturated velocity suddenly drops to a lower value. We can associate this sudden drop with the excitation of the internal breathing mode of the skyrmion. As can be seen in Fig.~\ref{FIG6}(b-C), a sizeable fraction of the skyrmion is located within the area of the applied RF field, which facilitates the excitation of internal modes.

Scenario D (red region) is the regime where the drive is so strong that the skyrmion will be destroyed shortly after the RF field is applied. In this limit, $H_{\text{a}}>3200$ mT, the skyrmion is driven so hard towards the edge that it is eventually pushed over the edge barrier. The topologically non-trivial skyrmion then unwinds and disappears. A snapshot of the destruction process is shown in Fig.~\ref{FIG6}(b-D).

\begin{figure}[t]
\centerline{\includegraphics[width=0.50\textwidth]{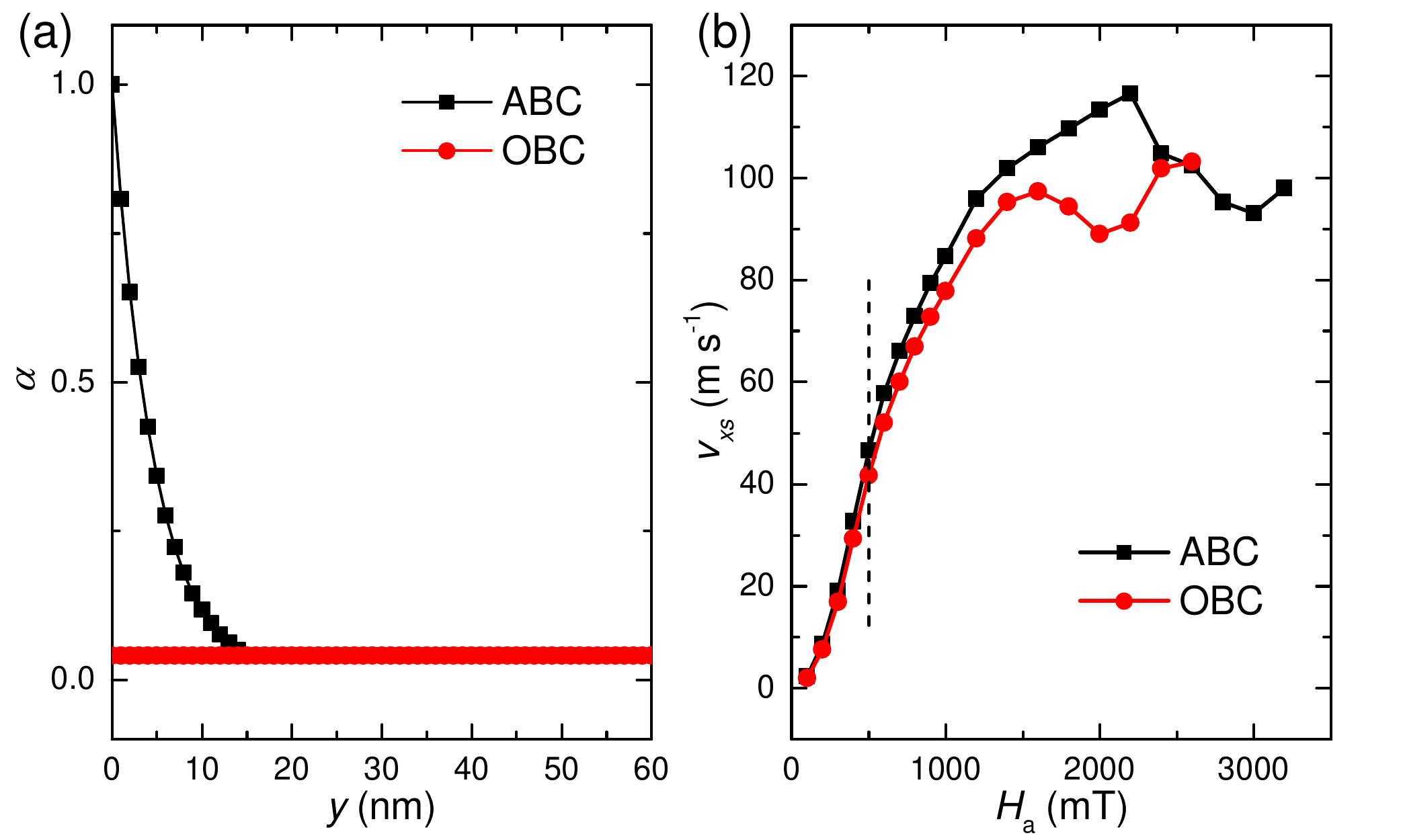}}
\caption{%
(a) Damping coefficient as a function of $y$ coordinate for the absorbing boundary conditions (ABCs) and open boundary conditions (OBCs).
(b) The saturation velocity $v_{xs}$ of a skyrmion driven by a transverse magnon current in a nanostrip as a function of the stimulating field amplitude $H_\text{a}$ for the models with ABCs and OBCs. The results for these two cases start to differ above $H=500$ mT (vertical dashed line).
}
\label{FIG7}
\end{figure}

It is worth mentioning that, for the purpose of avoiding the reflection of spin waves on the sample edges, absorbing boundary conditions (ABCs) are implemented in all simulations discussed above. However, in real-world experiments with nanoscale samples, the sample edges might indeed reflect spin waves. Hence, for comparison, we also performed simulations with open boundary conditions (OBCs) for the representative case of a $1000$ nm $\times$ $60$ nm $\times$ $1$ nm nanotrack, where the skyrmion is driven by a transverse magnon current. As shown in Fig.~\ref{FIG7}(a), the nanotrack with ABCs has an exponentially increasing damping coefficient at the edge, while the nanotrack with OBCs has a uniform damping coefficient. Fig.~\ref{FIG7}(b) shows the saturation velocity $v_{xs}$ of a skyrmion driven by a transverse magnon current as a function of the stimulating field amplitude $H_\text{a}$ for the models with ABCs and OBCs. The comparison shows that, when the stimulating field is relatively small ($H_\text{a}<500$ mT), the results obtained for the model with OBCs are in good agreement with those obtained for the model with ABCs. The reason is that the reflection of magnons is negligible if the exciting field is sufficiently small. Also, the skyrmion barely enters the region of increased damping at the edge. However, when the stimulating field is relatively large ($H_\text{a}>500$ mT), the results for $v_{xs}$ obtained for the model with OBCs differ quantitatively from those obtained for the model with ABCs, and the difference increases with increasing $H_\text{a}$. Nevertheless, it can be seen that qualitatively the four different scenarios can also be identified for the model with OBCs, which are now shifted to lower field values $H_\text{a}$.

\section{Summary}
\label{sec:summary}

We studied in depth the motion of an isolated magnetic skyrmion in a nanotrack driven by a magnonic momentum-transfer force provided by a spin wave. We considered two setups where spin waves are excited either longitudinal or transverse to the nanotrack. We find that the longitudinal driving is less efficient partly due to the damping of the spin waves.

For transverse driving, the skyrmion motion attains a steady-state with constant velocity along the track. We analyzed the saturated velocity, $v_{xs}$, as a function of the Gilbert damping $\alpha$ and the driving amplitude $H_{\text{a}}$ of the magnetic field that generates the spin waves. It obeys for low driving the scaling relation $v_{xs} \propto J/\alpha \propto |H_{\text{a}}|^2/\alpha$ where the magnon current amplitude $J \propto |H_{\text{a}}|^2$. The enhancement of the skyrmion velocity by the factor $1/\alpha$ is comparable to previous results obtained for the skyrmion motion in confined geometries but driven by electronic spin currents~\cite{Iwasaki_NNANO2013,Navau2016}.

For large transverse driving, the skyrmion is pushed into the driving layer and additional magnon modes and even the internal breathing mode of the skyrmion get excited. This limits a further increase of the skyrmion velocity and, finally, results in a destruction of the skyrmion.

Our study demonstrates that the position of a skyrmion can be efficiently manipulated by the magnonic momentum-transfer force, and it elucidates the principle of a magnon-driven skyrmion motion which might be of interest for practical applications.

\begin{acknowledgments}
X.Z. was supported by JSPS RONPAKU (Dissertation Ph.D.) Program. J.M. thanks Deutsche Telekom Stiftung, Bonn-Cologne Graduate School of Physics and Astronomy BCGS and the German funding program CRC 1238 for financial support. M.G. was supported by the SFB 1143 (Correlated Magnetism: From Frustration to Topology). Y.Z. acknowledges the support by the National Natural Science Foundation of China (Project No. 11574137) and Shenzhen Fundamental Research Fund under Grant No. JCYJ20160331164412545.
\end{acknowledgments}

\appendix
\section{MICROMAGNETIC MODELING}
\label{sec:methods_1}

We performed three-dimensional (3D) micromagnetic simulations by using the 1.2a5 release of the Object Oriented MicroMagnetic Framework (OOMMF) software developed at the National Institute of Standards and Technology (NIST)~\cite{OOMMF}. The simulations are handled by the OOMMF extensible solver (OXS) objects of the standard OOMMF distribution with the OXS extension module for simulating the interface-induced Dzyaloshinskii-Moriya interaction (DMI)~\cite{Rohart_PRB2013}.
The magnetization dynamics are controlled by the Landau-Lifshitz-Gilbert (LLG) equation~\cite{Gilbert_LLG,Landau_LLG,OOMMF} given as
\begin{equation}
\frac{d\boldsymbol{M}}{dt}=-\gamma_{\text{0}}\boldsymbol{M}\times\boldsymbol{H}_{\text{eff}}+\frac{\alpha}{M_{\text{S}}}(\boldsymbol{M}\times\frac{d\boldsymbol{M}}{dt}),
\label{LLG}
\end{equation}
where $\boldsymbol{M}$ is the magnetization, $\boldsymbol{H}_{\text{eff}}=-\mu_{0}^{-1}\frac{\partial E}{\partial \boldsymbol{M}}$ is the effective field, $t$ is the time, $\gamma_{\text{0}}$ is the gyromagnetic ratio, $\alpha$ is the Gilbert damping coefficient, and $M_{\text{S}}=|\boldsymbol{M}|$ is the saturation magnetization. The average energy density $E$ of the system contains the exchange energy, the anisotropy energy, the applied field (Zeeman) energy, the magnetostatic (demagnetization) energy, and the DMI energy terms, which is expressed as follows
\begin{align}
\label{EnergyDensity}
E =& \,A [\nabla(\frac{\boldsymbol{M}}{M_{\text{S}}})]^{2} - K \frac{(\boldsymbol{n}\cdot\boldsymbol{M})^{2}}{M_{\text{S}}^{2}} \\ \notag
& -\mu_{0} \boldsymbol{M}\cdot\boldsymbol{H} - \frac{\mu_{0}}{2} \boldsymbol{M}\cdot\boldsymbol{H}_{\text{d}}(\boldsymbol{M}) \\ \notag
& + \frac{D}{M_{\text{S}}^{2}} (M_{z}\frac{\partial M_{x}}{\partial x}+M_{z}\frac{\partial M_{y}}{\partial y}-M_{x}\frac{\partial M_{z}}{\partial x}-M_{y}\frac{\partial M_{z}}{\partial y}) \text{,}
\end{align}
where $A$ and $K$ are the exchange and anisotropy energy constants, respectively. $\boldsymbol{H}$ and $\boldsymbol{H}_{\text{d}}(\boldsymbol{M})$ are the applied field and the magnetostatic self-interaction field. The $M_x$, $M_y$ and $M_z$ are the components of the magnetization $\boldsymbol{M}$. The five terms at the right-hand side of Eq.~\ref{EnergyDensity} correspond to the exchange energy, the anisotropy energy, the applied field (Zeeman) energy, the magnetostatic (demagnetization) energy, and the DMI energy, respectively.
The material parameters used by the simulation program are adopted from Refs.~\onlinecite{Sampaio_NNANO2013,Yan_NCOMMS2014,Xichao_SREP2015A,Xichao_SREP2015B,Xichao_SREP2015C,Xichao_NANOTECH2015,Yan_NCOMMS2015,Xichao_NCOMMS2016}: the Gilbert damping coefficient $\alpha=0.01\sim 0.05$, the gyromagnetic ratio $\gamma=-2.211\times 10^{5}$ m A$^{-1}$ s$^{-1}$, the saturation magnetization $M_{\text{S}}=580$ kA m$^{-1}$, the exchange stiffness $J=15$ pJ m$^{-1}$, the interface-induced DMI constant $D=3.5$ mJ m$^{-2}$, and the perpendicular magnetic anisotropy (PMA) constant $K=0.8$ MJ m$^{-3}$, unless otherwise specified. The simulated models are discretized into tetragonal cells with the optimum cell size of $2$ nm $\times$ $2$ nm $\times$ $1$ nm, which offers a good trade-off between the computational accuracy and efficiency. Absorbing boundary conditions are implemented in all simulations (except the simulations given in Fig.~\ref{FIG7}) in order to avoid the reflection of spin waves on the sample edges.




\begin{thebibliography}{99}

\bibitem{Roszler_NATURE2006} Roszler, U. K., Bogdanov, A. N. \& Pfleiderer, C. Spontaneous skyrmion ground states in magnetic metals. \textit{Nature} \textbf{442}, 797-801 (2006).

\bibitem{Nagaosa_NNANO2013} Nagaosa, N. \& Tokura, Y. Topological properties and dynamics of magnetic skyrmions. \textit{Nat. Nano.} \textbf{8}, 899-911 (2013).

\bibitem{Seki_BOOK2016} Seki, S. \& Mochizuki, M. \textit{Skyrmions in Magnetic Materials} (Springer, 2016).

\bibitem{Muhlbauer_SCIENCE2009} M\"{u}hlbauer, S. \textit{et al.} Skyrmion lattice in a chiral magnet. \textit{Science} \textbf{323}, 915-919 (2009).

\bibitem{Yu_NATURE2010} Yu, X. Z. \textit{et al.} Real-space observation of a two-dimensional skyrmion crystal. \textit{Nature} \textbf{465}, 901-904 (2010).

\bibitem{Heinze_NPHYS2011} Heinze, S. \textit{et al.} Spontaneous atomic-scale magnetic skyrmion lattice in two dimensions. \textit{Nat. Phys.} \textbf{7}, 713-718 (2011).

\bibitem{Schulz_NPHYS2012} Schulz, T. \textit{et al.} Emergent electrodynamics of skyrmions in a chiral magnet. \textit{Nat. Phys.} \textbf{8}, 301-304 (2012).

\bibitem{Romming_SCIENCE2013} Romming, N. \textit{et al.} Writing and deleting single magnetic skyrmions. \textit{Science} \textbf{341}, 636-639 (2013).

\bibitem{Finazzi_PRL2013} Finazzi, M. \textit{et al.} Laser-induced magnetic nanostructures with tunable topological properties. \textit{Phys. Rev. Lett.} \textbf{110}, 177205 (2013).

\bibitem{Kezsmarki_NMATER2015} K{\'e}zsm{\'a}rki, I. \textit{et al.} N\'eel-type skyrmion lattice with confined orientation in the polar magnetic semiconductor GaV$_{4}$S$_8$. \textit{Nat. Mater.} \textbf{14}, 1116-1122 (2015).

\bibitem{Schwarze_NMATER2015} Schwarze, T. \textit{et al.} Universal helimagnon and skyrmion excitations in metallic, semiconducting and insulating chiral magnets. \textit{Nat. Mater.} \textbf{14}, 478-483 (2015).

\bibitem{Du_NCOMMS2015} Du, H. F. \textit{et al.} Edge-mediated skyrmion chain and its collective dynamics in a confined geometry. \textit{Nat. Commun.} \textbf{6}, 8504 (2015).

\bibitem{Nii_NCOMMS2015} Nii, Y. \textit{et al.} Uniaxial stress control of skyrmion phase. \textit{Nat. Commun.} \textbf{6}, 8539 (2015).

\bibitem{Wanjun_SCIENCE2015} Jiang, W. \textit{et al.} Blowing magnetic skyrmion bubbles. \textit{Science} \textbf{349}, 283-286 (2015).

\bibitem{Boulle_NNANO2016} Boulle, O. \textit{et al.} Room-temperature chiral magnetic skyrmions in ultrathin magnetic nanostructures. \textit{Nat. Nano.} \textbf{11}, 449-454 (2016).

\bibitem{Wanjun_NPHYS2017} Jiang, W. \textit{et al.} Direct observation of the skyrmion Hall effect. \textit{Nat. Phys.} \textbf{13}, 162-169 (2017).

\bibitem{Wanjun_AIP2016} Jiang, W. \textit{et al.} Mobile N\'eel skyrmions at room temperature: status and future. \textit{AIP Adv.} \textbf{6}, 055602 (2016)

\bibitem{WOO_NMAT2016} Woo, S. \textit{et al.} Observation of room-temperature magnetic skyrmions and their current-driven dynamics in ultrathin metallic ferromagnets. \textit{Nat. Mater.} \textbf{15}, 501-506 (2016).

\bibitem{Fert_NNANO2013} Fert, A.,  Cros, V. \& Sampaio, J. Skyrmions on the track. \textit{Nat. Nano.} \textbf{8}, 152-156 (2013).

\bibitem{Mueller_ARXIV2016} M\"uller, J. Magnetic Skyrmions on a two-lane racetrack. \textit{New J. Phys.} \textbf{19} 025002 (2017).

\bibitem{Sampaio_NNANO2013} Sampaio, J., Cros, V., Rohart, S., Thiaville, A. \& Fert, A. Nucleation, stability and current-induced motion of isolated magnetic skyrmions in nanostructures. \textit{Nat. Nano.} \textbf{8}, 839-844 (2013).

\bibitem{Iwasaki_NNANO2013} Iwasaki, J., Mochizuki, M. \& Nagaosa, N. Current-induced skyrmion dynamics in constricted geometries. \textit{Nat. Nano.} \textbf{8}, 742-747 (2013).

\bibitem{Sun_PRL2013} Sun, L. \textit{et al.} Creating an artificial two-dimensional skyrmion crystal by nanopatterning. \textit{Phys. Rev. Lett.} \textbf{110}, 167201 (2013).

\bibitem{Tomasello_SREP2014} Tomasello, R. \textit{et al.} A strategy for the design of skyrmion racetrack memories. \textit{Sci. Rep.} \textbf{4}, 6784 (2014).

\bibitem{Yan_NCOMMS2014} Zhou, Y. \& Ezawa, M. A reversible conversion between a skyrmion and a domain-wall pair in a junction geometry. \textit{Nat. Commun.} \textbf{5}, 4652 (2014).

\bibitem{Xichao_SREP2015A} Zhang, X. \textit{et al.} Skyrmion-skyrmion and skyrmion-edge repulsions in skyrmion-based racetrack memory. \textit{Sci. Rep.} \textbf{5}, 7643 (2015).

\bibitem{Xichao_SREP2015B} Zhang, X., Ezawa, M. \& Zhou, Y. Magnetic skyrmion logic gates: conversion, duplication and merging of skyrmions. \textit{Sci. Rep.} \textbf{5}, 9400 (2015).

\bibitem{Xichao_SREP2015C} Zhang, X., Zhou, Y., Ezawa, M., Zhao, G. P. \&  Zhao, W. Magnetic skyrmion transistor: skyrmion motion in a voltage-gated nanotrack. \textit{Sci. Rep.} \textbf{5}, 11369 (2015).

\bibitem{Yan_NCOMMS2015} Zhou, Y. \textit{et al.} Dynamically stabilized magnetic skyrmions. \textit{Nat. Commun.} \textbf{6}, 8193 (2015).

\bibitem{Fusheng_NANOLETT2015} Ma, F., Zhou, Y., Braun, H. B. \& Lew, W. S. Skyrmion-based dynamic magnonic crystal. \textit{Nano Lett.} \textbf{15}, 4029-4036 (2015).

\bibitem{Beg_SREP2015} Beg, M. \textit{et al.} Ground state search, hysteretic behaviour, and reversal mechanism of skyrmionic textures in confined helimagnetic nanostructures. \textit{Sci. Rep.} \textbf{5}, 17137 (2015).

\bibitem{Upadhyaya_PRB2015} Upadhyaya, P., Yu, G. P., Amiri, P. K. \& Wang, K. L. Electric-field guiding of magnetic skyrmions. \textit{Phys. Rev. B} \textbf{92}, 134411 (2015).

\bibitem{Koshibae_JJAP2015} Koshibae, W. \textit{et al.} Memory functions of magnetic skyrmions. \textit{Japan. J. Appl. Phys.} \textbf{54}, 053001 (2015).

\bibitem{Xichao_NCOMMS2016} Zhang, X., Zhou, Y. \& Ezawa, M. Magnetic bilayer-skyrmions without skyrmion hall effect. \textit{Nat. Commun.} \textbf{7}, 10293 (2016).

\bibitem{Wang_SREP2016} Yuan, H. Y. \& Wang, X. R. Skyrmion creation and manipulation by nano-second current pulses. \textit{Sci. Rep.} \textbf{6}, 22638 (2016).

\bibitem{Senfu_NJP2015} Zhang, S. \textit{et al.} Current-induced magnetic skyrmions oscillator. \textit{New J. Phys.} \textbf{17} 023061 (2015).

\bibitem{Beg_PRB2017} Beg, M. \textit{et al.} Dynamics of skyrmionic states in confined helimagnetic nanostructures. \textit{Phys. Rev. B} \textbf{95}, 014433 (2017).

\bibitem{Bazeia_JMMM2017} Bazeia, D., Ramos, J. \& Rodrigues, E. Topological strength of magnetic skyrmions. \textit{J. Magn. Magn. Mater.} \textbf{423} 411-420 (2017).

\bibitem{Lin_JAP2014} Lin, S.-Z., Reichhardt, C., Bastisa, C.D. \& Saxena, A. Dynamics of skyrmions in chiral magnets: dynamic phase transitions and equation of motion, \textit{J. Appl. Phys.} \textbf{115}, 17D109 (2014).

\bibitem{Reichhardt_PRB2015} Reichhardt, C. \& Olson Reichhardt, C.J. Shapiro steps for skyrmion motion on a washboard potential with longitudinal and transverse ac drives, \textit{Phys. Rev. B} \textbf{92}, 224432 (2015).

\bibitem{Guoqiang_NL2017} Yu, G. \textit{et al.} Room-temperature skyrmion shift device for memory application, \textit{Nano Lett.} \textbf{17}, 261-268 (2017).

\bibitem{Schutte_PRB2014A} Sch\"utte, C. \& Garst, M. Magnon-skyrmion scattering in chiral magnets. \textit{Phys. Rev. B} \textbf{90}, 094423 (2014).

\bibitem{Schutte_PRB2014B} Sch\"utte, C., Iwasaki, J., Rosch, A. \& Nagaosa, N. Inertia, diffusion, and dynamics of a driven skyrmion. \textit{Phys. Rev. B} \textbf{90}, 174434 (2014).

\bibitem{Iwasaki_PRB2014} Iwasaki, J., Beekman, A. J. \& Nagaosa, N. Theory of magnon-skyrmion scattering in chiral magnets. \textit{Phys. Rev. B} \textbf{89}, 064412 (2014).

\bibitem{Schroeter_LTP2015} Schroeter, S. \& Garst, M. Scattering of high-energy magnons off a magnetic skyrmion. \textit{Low Temp. Phys.} \textbf{41}, 817 (2015).

\bibitem{Xichao_NANOTECH2015} Zhang, X. \textit{et al.} All-magnetic control of skyrmions in nanowires by a spin wave. \textit{Nanotechnology} \textbf{26}, 225701 (2015).

\bibitem{Kim_ARXIV2015} Kim, J. \& Kim, S. K. Origin of robust interaction of spin waves with a single skyrmion in perpendicularly magnetized nanostripes. Preprint at http://arxiv.org/abs/1508.05682 (2015). (Accessed: 25$^{\text{th}}$ August 2015)

\bibitem{Wanjun_PRL2013} Jiang, W. \textit{et al.} Direct imaging of thermally driven domain wall motion in magnetic insulators. \textit{Phys. Rev. Lett.} \textbf{110}, 177202 (2013).

\bibitem{Wang_PRL2011} Yan, P., Wang, X. S. \& Wang, X. R. All-magnonic spin-transfer torque and domain wall propagation. \textit{Phys. Rev. Lett.} \textbf{107}, 177207 (2011).

\bibitem{Wang_PRL2012} Wang, X. S., Yan, P., Shen, Y. H., Bauer, G. E. W. \& Wang, X. R. Domain wall propagation through spin wave emission. \textit{Phys. Rev. Lett.} \textbf{109}, 167209 (2012).

\bibitem{Thiele_PRL1973} Thiele, A. A. Steady-state motion of magnetic domains. \textit{Phys. Rev. Lett.} \textbf{30}, 230-233 (1973).

\bibitem{Rohart_PRB2013} Rohart, S. \& Thiaville, A. Skyrmion confinement in ultrathin film nanostructures in the presence of Dzyaloshinskii-Moriya interaction. \textit{Phys. Rev. B} \textbf{88}, 184422 (2013).

\bibitem{Navau2016} Navau, C., Del-Valle, N. \& Sanchez, A., Analytical trajectories of skyrmions in confined geometries: Skyrmionic racetracks and nano-oscillators, \textit{Phys. Rev. B} \textbf{94}, 184104 (2016).

\bibitem{OOMMF} Donahue, M. J. \& Porter, D. G. OOMMF user's guide, version 1.0. \textit{Interagency Report} \textbf{NISTIR 6376} (1999).

\bibitem{Gilbert_LLG} Gilbert, T. L. A Lagrangian formulation of the gyromagnetic equation of the magnetization field. \textit{Phys. Rev.} \textbf{100}, 1243 (1955).

\bibitem{Landau_LLG} Landau, L. \& Lifshitz, E. On the theory of the dispersion of magnetic permeability in ferromagnetic bodies. \textit{Phys. Zeitsch. der Sow.} \textbf{8}, 153-169 (1935).

\end{thebibliography}
\end{document}